\documentclass[aps,prl,twocolumn,superscriptaddress]{revtex4-1}

\usepackage{advdate}
\usepackage{autobreak}
\usepackage{amsmath,amssymb}
\usepackage{braket}
\usepackage{mathtools}
\usepackage{amsfonts}
\usepackage{bm}
\usepackage{ascmac}
\usepackage{graphicx}
\usepackage{color}
\usepackage{floatrow}
\usepackage{here}
\usepackage{multirow}
\usepackage{afterpage}
\newcommand{\Z}{$\mathbb{Z}$}
\newcommand{\Zt}{$\mathbb{Z}_{2}$}

\begin{document}

\title{Universal platform of point-gap topological phases from topological materials}

\author{Daichi Nakamura}
\email{daichi.nakamura@yukawa.kyoto-u.ac.jp}
\affiliation{Center for Gravitational Physics and Quantum Information, Yukawa Institute for Theoretical Physics, Kyoto University, Kyoto 606-8502, Japan}

\author{Kazuya Inaka}
\email{kazuya.inaka@yukawa.kyoto-u.ac.jp}
\affiliation{Center for Gravitational Physics and Quantum Information, Yukawa Institute for Theoretical Physics, Kyoto University, Kyoto 606-8502, Japan}

\author{Nobuyuki Okuma}
\email{okuma@hosi.phys.s.u-tokyo.ac.jp}
\affiliation{Graduate School of Engineering, Kyushu Institute of Technology, Kitakyushu 804-8550, Japan}

\author{Masatoshi Sato}
\email{msato@yukawa.kyoto-u.ac.jp}
\affiliation{Center for Gravitational Physics and Quantum Information, Yukawa Institute for Theoretical Physics, Kyoto University, Kyoto 606-8502, Japan}

\begin{abstract}
Whereas point-gap topological phases are responsible for exceptional phenomena intrinsic to non-Hermitian systems, 
their realization in quantum  materials is still elusive.
Here we propose a simple and universal platform of point-gap topological phases constructed from Hermitian topological insulators and superconductors.
We show that $(d-1)$-dimensional point-gap topological phases are realized by making a boundary in $d$-dimensional topological insulators and superconductors dissipative.
A crucial observation of the proposal is that adding a decay constant to boundary modes in $d$-dimensional topological insulators and superconductors is topologically equivalent to attaching a $(d-1)$-dimensional point-gap topological phase to the boundary.
We furthermore establish the proposal from the extended version of the Nielsen-Ninomiya theorem, relating dissipative gapless modes to point-gap topological numbers. 
From the bulk-boundary correspondence of the point-gap topological phases, the resultant point-gap topological phases exhibit exceptional boundary states or in-gap higher-order non-Hermitian skin effects.  
\end{abstract}

\maketitle

{\it Introduction.--}
There is increasing interest in the studies of non-Hermitian physics \cite{Bender07, MB04, Moiseyev11}. 
Among them, a recent trend is the study of topology in non-Hermitian systems \cite{AGU20,BBK21,OS23}.
The prime motivation for such a research direction is that both non-Hermitian and topological systems exhibit characteristic boundary phenomena \cite{RL09,HH11,ESHK11,Schomerus13,SZF18,YW18,KEBB18}.
Certain non-Hermitian systems show a boundary phenomenon called the non-Hermitian skin effect \cite{YW18,YM19}, where a macroscopic number of bulk states are localized at the boundary. 
On the other hand, the bulk-boundary correspondence \cite{Hatsugai93}, in which bulk topological invariants count the number of gapless boundary states, is one of the most notable concepts in topological systems \cite{HK10,QZ11}.
In topological systems with the non-Hermiticity, both the non-Hermitian skin effect and the topological boundary states can coexist, and the bulk-boundary correspondence should hold in an unconventional manner \cite{YW18,KEBB18,YSW18,HBR19,SYW19,NBS22}.

The generalization of the topological classification to non-Hermitian systems is also of interest.  In the original classification of Hermitian topological insulators and superconductors \cite{SRFL08,Kitaev09,RSFL10,CTSR16}, the gapped topology is mathematically characterized by the absence of the energy eigenstates of the Hamiltonian at the Fermi energy, $E\neq E_{\rm F}$. The natural extension to non-Hermitian systems is real line-gap topology defined by $\mathrm{Re}(E-E_ {\rm F})\neq0$ \cite{ESHK11, KSUS19}. 
%The real line-gap topology describes the competition between the non-Hermiticity and Hermitian-like topology, as introduced above.
Mathematically, the real line-gapped Hamiltonians are smoothly deformed into Hermitian-gapped Hamiltonians without closing the real-line gap \cite{ESHK11,KSUS19,AGU20}. Therefore, the physical consequence of the real line-gapped topology is the bulk-boundary correspondence, as in the case of the Hermitian topological phases. More generally, the line-gapped spectrum is defined as a spectrum that does not cross a specific line in the complex plane \cite{KSUS19}. For instance, if one chooses the real axis in the complex energy plane as the reference line, such a spectrum defines the imaginary line-gap topology, which is adiabatically connected to anti-Hermitian topological phases \cite{KSUS19}. 

Remarkably, the non-Hermiticity enables another extension of topology, the point-gap topology defined by $E\neq E_{\rm P}$ \cite{GAKTHU18,KSUS19}. 
Typically, a spectrum with nontrivial point-gap topology surrounds the reference point $E_{\rm P}$ in the complex energy plane, and thus the point-gap topology is distinct from any Hermitian-like topology. 
The topological classification of point-gapped Hamiltonians has been established \cite{ZL19,KSUS19}, and the physical consequences of the point-gap topological phases have been explored \cite{GAKTHU18,BKS19,OS19,OKSS20,ZYF20,OS21,KSR21,YSHC21,VDNS21,DSSFTBN21,HZLW22,NBS22,DSSFTBN21,LAZV19,TK20,BS21,LLG19,LZ19,LLG20,OTY20,KSS20,FHW21,SO21,PTGPSG21,KP21,ZTJLC21,GLSH21,GLS21,ZCHBLSZ21,FKHBF21,LLWLL22,ZG22,MF22,HI22,SLSHZZSGLT22,WJ23,RBGUCCC23,NONSS23}.
In particular, it has been shown that the non-Hermitian skin effects originate from one-dimensional point-gap topological numbers, {\it i.e.} the spectral winding number \cite{OKSS20,ZYF20} or the $\mathbb{Z}_2$ number \cite{OKSS20}.
Also, higher-dimensional point-gap topological phases may support non-Hermitian skin modes localized at topological defects, mimicking the anomaly-induced catastrophes \cite{OKSS20,OS21,KSR21,MF22,OS23}.
Depending on the dimension and symmetry of the system, higher-dimensional point-gap topological phases may also host boundary modes \cite{YSHC21,VDNS21,DSSFTBN21,HZLW22,NBS22,MF22}, and one of such point-gap topological phases is called an exceptional topological insulator \cite{DSSFTBN21}. 
Furthermore, for the fundamental symmetry classes called AZ$^\dagger$ classes (see below), the correspondence between $d$-dimensional point-gap topological phases and $(d-1)$-dimensional anomalous gapless modes was suggested \cite{LAZV19}, and later proved as the extended Nielsen-Ninomiya theorem \cite{BS21}.

%While there is no unified interpretation applicable to all the fundamental symmetry classes, the non-Hermitian skin effect and its generalization for some symmetry classes and dimensions, called symmetry-protected/higher-dimensional skin effects, can be characterized as point-gapped topological phenomena.
%\textcolor{red}{anomaly interpretation to Nielsen Ninomiya nado ha omakase simasu. skin effect to anomaly interpretation mo ichiou genkyuu shite kudasai...}

%(General introduction for non-Hermitian topological phases. Please write this part, Okuma-san)

\begin{figure}[btp]
 \begin{center}
  \includegraphics[scale=0.12]{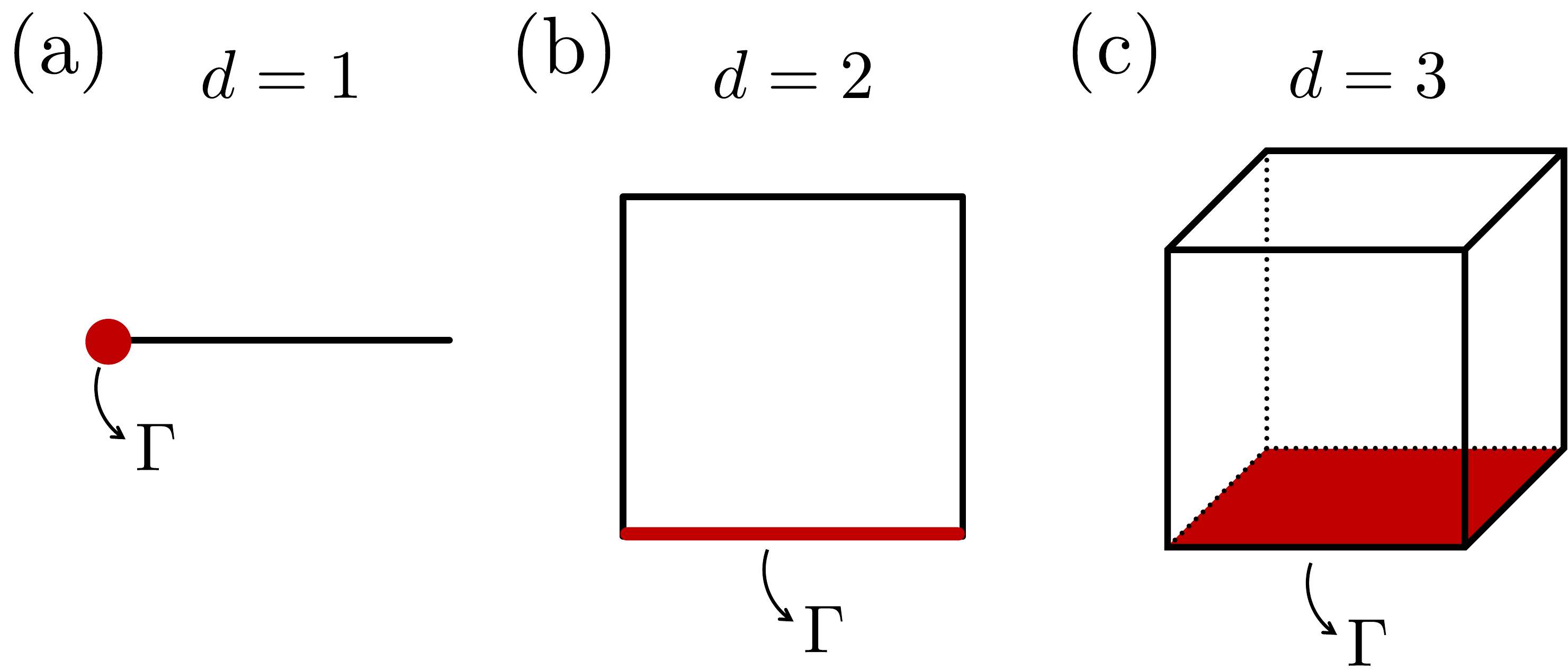}
  \caption{Universal plat forms of point-gap topological phases. $d$-dimensional topological insulators and superconductors are coupled to the environment at $x_d=1$.  \label{fig:plat_forms}}
  \end{center}
\end{figure}

Whereas point-gap topological phases are responsible for these exceptional phenomena intrinsic to non-Hermitian systems, 
their realization in quantum  materials is still elusive.
This paper proposes a simple and universal platform of point-gap topological phases in quantum materials.
As illustrated in Fig.\ref{fig:plat_forms}, the platform consists of a $d$-dimensional topological insulator or superconductor where one of the boundaries is coupled to the environment and thus dissipative. 
We show that the dissipation-induced decay constant of the topological boundary modes results in a $(d-1)$-dimensional non-trivial point-gap topological number, {\it i.e.} a $(d-1)$-dimensional point-gap topological phase.
We also predict exceptional boundary states or in-gap higher-order non-Hermitian skin effects
based on the bulk-boundary correspondence for point-gap topological phases \cite{DSSFTBN21,NBS22}.

\begin{figure}[btp]
 \begin{center}
  \includegraphics[scale=0.12]{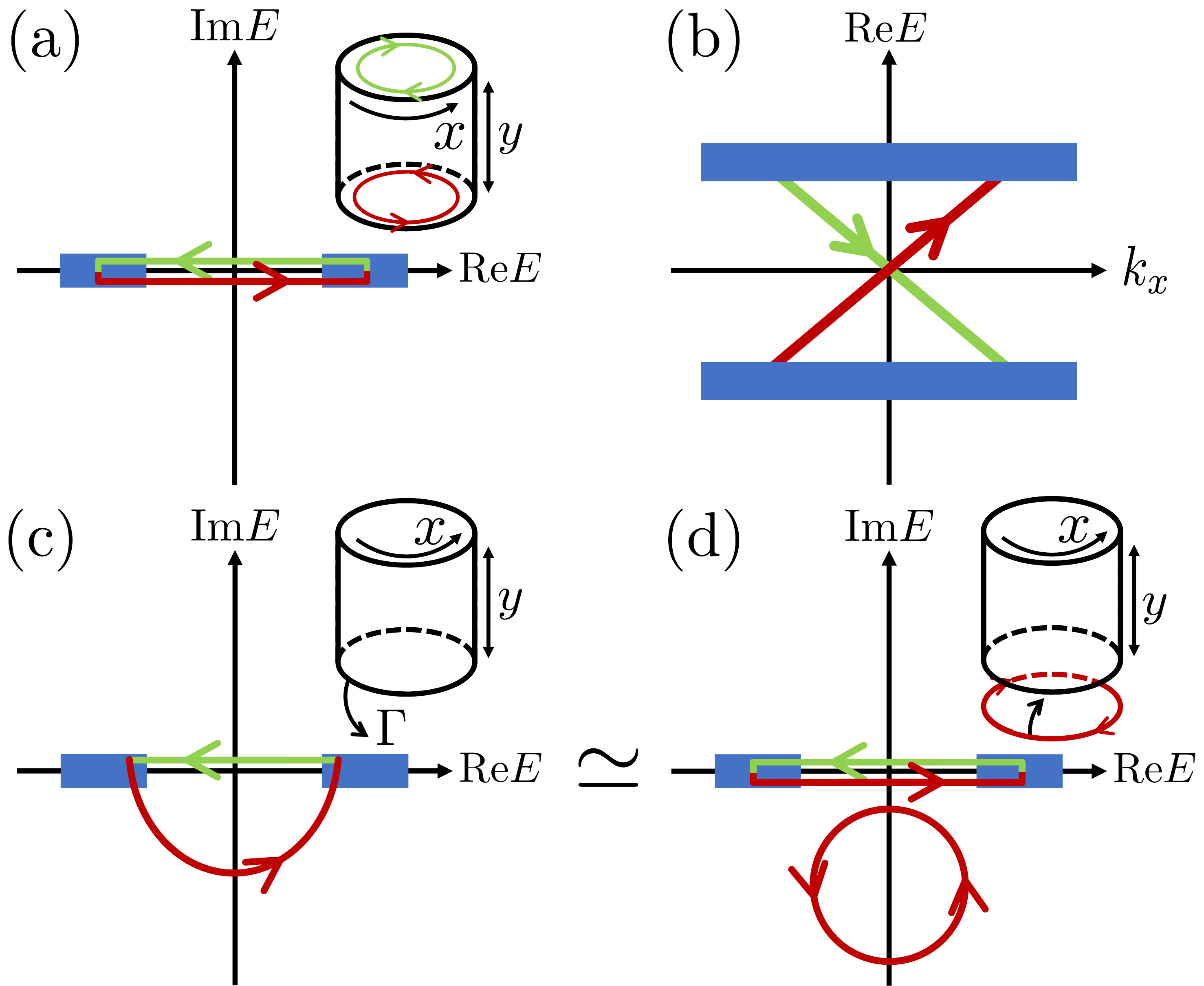}
  \caption{(a, b) Chern insulator under the periodic boundary condition in the $x$-direction and the open boundary condition in the $y$-direction. A chiral edge mode (red) at the bottom boundary and an anti-chiral edge mode (green) at the top boundary form a loop on the real axis in the complex energy plane.  (c) Chern insulator (a) with the decay constant term $\Gamma$ at the bottom boundary. (d) Chern insulator (a) attached the Hatano-Nelson model at the bottom boundary.\label{fig:chiral_edge}}
  \end{center}
\end{figure}

{\it Non-trivial topology from decay constant.--}
Let us start with a Chern insulator with the periodic boundary condition in the $x$-direction and the open boundary conditions at $y=1, L_y$ in the $y$-direction.
See Fig.\ref{fig:chiral_edge} (a).
The system supports a chiral edge mode at $y=1$ and an anti-chiral edge mode at $y=L_y$. 
If we couple one of the open boundaries, say $y=1$, to the environment, the chiral edge mode at $y=1$ gets the decay constant in addition to the linear spectrum, 
\begin{align}
h_{\rm edge}(k_x)=v k_x-i\Gamma,    
\label{eq:chiral_edge}
\end{align}
where $k_x$ is the momentum in the $x$-direction, $v(>0)$ is the group velocity,  and $\Gamma(>0)$ is the decay constant.  
At first sight, the decay constant seems not to change the topology of the system, but it does change, as we see below.  

To see the hidden topology due to the decay constant, 
we show that the complex spectrum in Eq.(\ref{eq:chiral_edge}) is equivalently obtained by attaching a one-dimensional point-gap topological phase to the original Hermitian chiral edge state:
The effective Hamiltonian of the attached system is
\begin{align}
h_{\rm attach}(k_x)=\left(
\begin{array}{cc}
 vk_x    & \Delta \\
\Delta^*   & -v\sin k_x+i(\Gamma/2)\cos k_x-i(\Gamma/2)
\end{array}
\right),
\label{eq:attach}
\end{align}
where the diagonal components describe the chiral edge state and the one-dimensional point-gap topological phase, respectively, and $\Delta$ $(|\Delta|\ll 1)$ is the coupling between them.
The one-dimensional point-gap topological phase is the tight-binding lattice model with asymmetric hoppings (Hatano-Nelson model \cite{HN96,HN97,HN98}) and supports a non-zero spectral winding number. 
Remarkably, around $k_x= 0$, the attached system shows the spectrum
$
E_0(k_x)=\pm \sqrt{(vk_x)^2+|\Delta|^2},    
$
and thus the original edge state has a gap in the real part of the spectrum, but there appears another chiral mode around $k_x=\pi$, 
$
E_\pi(k_x)=v (k_x-\pi)-i\Gamma.    
$
Therefore, by shifting the origin of the momentum space, the attached system reproduces the complex spectrum in Eq.(\ref{eq:chiral_edge}).
Since the attached system in Eq. (\ref{eq:attach}) has a non-zero spectral winding number, the dissipative chiral edge state in Eq.(\ref{eq:chiral_edge}) also should have the same non-zero winding number. 

\begin{figure}[btp]
 \begin{center}
  \includegraphics[scale=0.13]{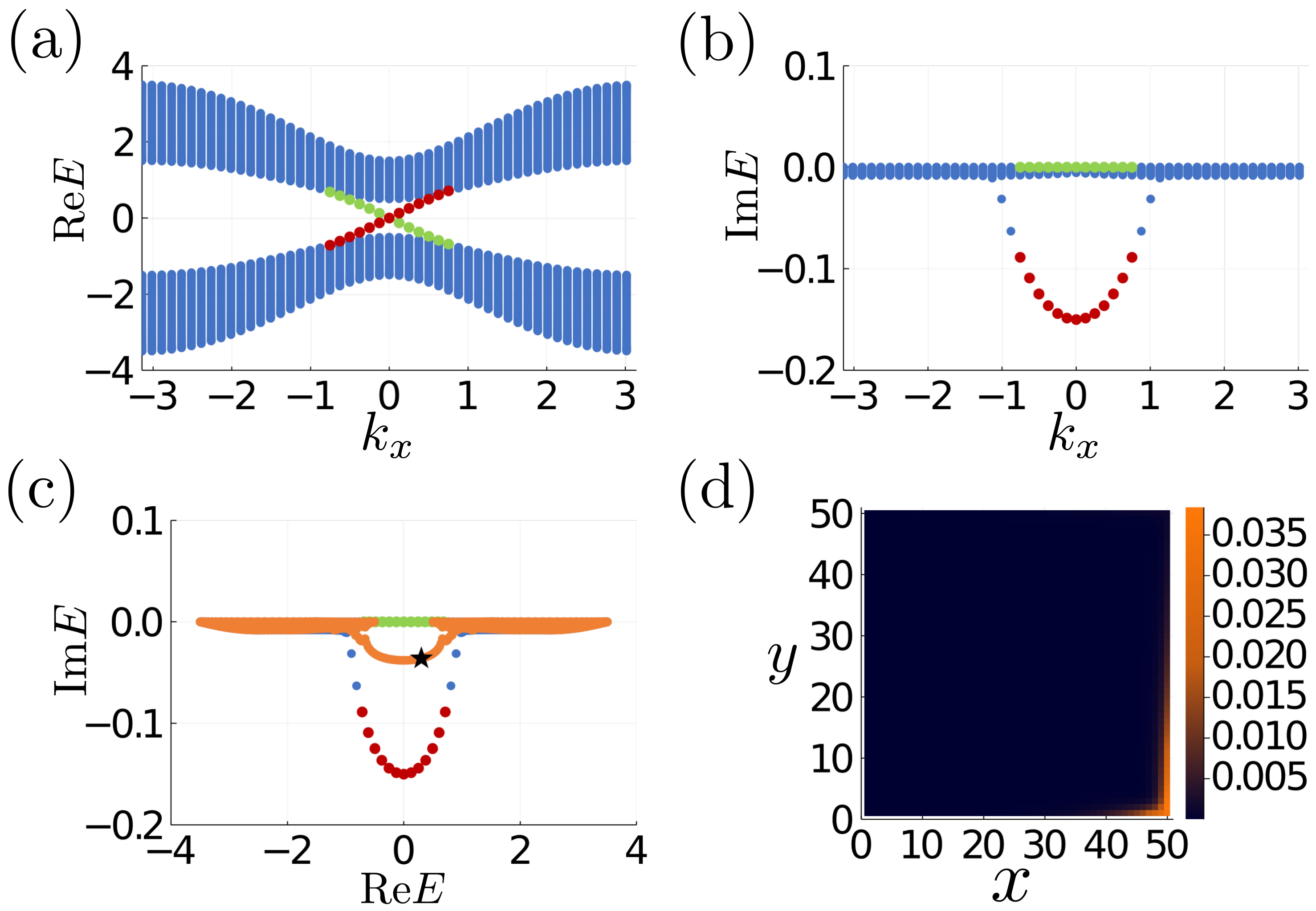}
  \caption{The QWZ model ($m=-1.5$, $L_x=L_y=50$) in Eq.(\ref{eq:QWZ}) with $-i\Gamma \sigma_0$ term ($\Gamma=0.2$) at the $y=1$ boundary. (a, b) The real and imaginary parts of the spectrum under the periodic boundary condition in the $x$-direction. (c) Comparison of the complex energy spectra under the open boundary condition (orange) and the periodic boundary condition (other colors) in the $x$-direction. (d) The skin mode with the energy $E=0.30-0.04 i$ (the star symbol in (c)).
  %$E=0.30283070292925296 - 0.03596762268299295i$ 
  \label{fig:QWZ}}
  \end{center}
\end{figure}

\begin{figure}[btp]
 \begin{center}
  \includegraphics[scale=0.13]{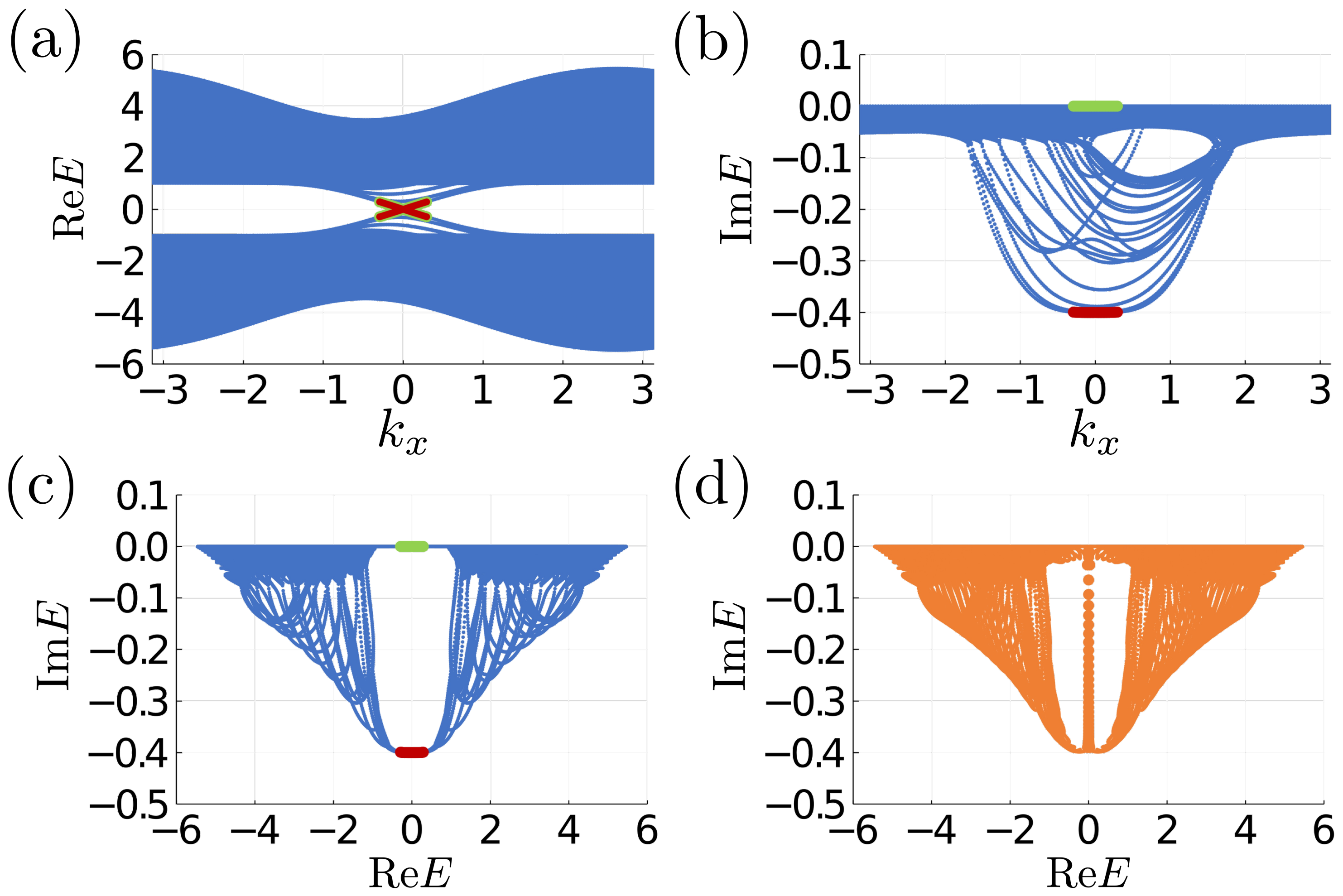}
  \caption{ 3D chiral symmetric topological insulator  ($L_x=300$, $L_y=L_z=20$) in Eq.(\ref{eq:3DCSTI}) with 
  $-i\Gamma\sigma_0\tau_0$ ($\Gamma=0.2$) term at the $z=1$ boundary. (a,b,c) The real and imaginary parts of the spectrum under the periodic boundary conditions in the $x$ and $y$-directions. (d) The complex spectrum under the periodic boundary condition in the $x$-direction and the open boundary condition in the $y$-direction. There appear in-gap boundary modes. 
 \label{fig:3dCSTI}}
  \end{center}
\end{figure}

Whereas the above argument is rather heuristic, we also have a convincing discussion on the non-trivial topology:  
For a rigorous discussion, we assume that the decay constant induced on the boundary is uniform and thus retains the lattice translation symmetry along the edge (namely the $x$-direction) \footnote{For a microscopic derivation of such a dissipation, see Sec.S2 in Supplement Material, which includes Refs. \cite{Datta95,ATEKOW14}.}.  
Then, from the Bloch theorem, any energy eigenstate in the present model should be labeled by $k_x$, and we have the $2\pi$-periodicity in $k_x$ for the energy eigenstates.
This means that the chiral edge state at $y=1$ and the anti-chiral edge one at $y=L_y$ should be exchanged when changing $k_x$ by $2\pi$ since they cannot go back to themself after the one-period.
As a result, they make a loop in the complex energy plane, as illustrated in Fig.\ref{fig:chiral_edge}.
In the absence of dissipation, the loop sticks to the real axis of the complex energy plane, as in Fig.\ref{fig:chiral_edge} (a), which implies that the spectral winding number is zero.
However, once the chiral edge mode at $y=1$ has the imaginary part of the energy due to dissipation, 
the loop is extended to the imaginary energy direction so it immediately gets a non-zero spectral winding number as shown in Fig.\ref{fig:chiral_edge} (c).
Therefore, the decay constant of the chiral edge mode results in the non-trivial point-gap topology.
We can also confirm that the decay of the chiral edge mode is topologically the same as the attachment of the one-dimensional Hatano-Nelson model to the boundary:
The Hatano-Nelson model gives a loop spectrum in the complex energy plane in Fig.\ref{fig:chiral_edge} (d). Then through the reconnection of the spectra between the Hatano-Nelson model and the chiral edge mode, we smoothly obtain the complex spectrum in Fig.\ref{fig:chiral_edge} (c) \footnote{See Sec.S4 in Supplement Material for a more detailed relation between the Hatano-Nelson model and the chiral edge state with dissipation.}. 

We can make a concrete prediction due to the non-trivial spectral winding number from the decay constant.    
Namely, the system exhibits the non-Hermitian skin effect under the open boundary condition in the $x$-direction.
To check the prediction, we consider the Chern insulator modeled by Qi, Wu, and Zhang (QWZ) \cite{QWZ06}
\begin{align}
H_{\rm QWZ}({\bm k})&= \sin k_x \sigma_x
+\sin k_y \sigma_y
\nonumber\\
&+(m+\cos k_x+\cos k_y)\sigma_z,   
\label{eq:QWZ}
\end{align}
which has the Chern number 1 for $-2<m<0$. Here $\sigma_{\mu=0,x,y,z}$ are the Pauli matrices.
When we impose the open boundary condition on the $y$-direction and put the imaginary onsite potential $-i\Gamma\sigma_0$ along the boundary at $y=1$, the chiral edge state gets a finite lifetime, as shown in Figs.~\ref{fig:QWZ}(a) and \ref{fig:QWZ}(b).
Then, if we further impose the open boundary condition on the $x$-direction, the system shows the non-Hermitian skin effect. See Fig.\ref{fig:QWZ}(c). We have $O(L)$ corner skin modes shown in Fig.\ref{fig:QWZ}(d).
This behavior entirely agrees with the prediction \footnote{See Sec.S1 in Supplement Material for spatial profiles of open boundary modes, which includes Refs. \cite{OS23,NONSS23}.}.

{\it Universal platform of point-gap topological phases.--}
So far, we have considered a two-dimensional Chern insulator and have shown how to realize a one-dimensional point-gap topological phase from the Chern insulator.  
Now we generalize the idea to other topological insulators and superconductors.
Let us consider a $d$-dimensional topological insulator or superconductor.  
Under the open boundary conditions at $x_d=1, L_d$ in the $x_d$-direction, we have topological gapless boundary modes with an opposite topological charge at opposite boundaries. 
Then, if we couple one of the boundaries, say $x_d=1$, to the environment, the topological gapless modes at $x_d=1$ get a finite lifetime, namely the imaginary part of the spectrum, due to dissipation.
As we will see shortly, this configuration realizes a $(d-1)$-dimensional point-gap topological phase. 

To prove the above statement, we first clarify the symmetry of the system. 
The fundamental onsite symmetries for topological insulators and superconductors are time-reversal (TRS), particle-hole (PHS), and their combination, chiral symmetry (CS).
These Altland-Zirnbauer (AZ) symmetries \cite{AZ97} protect topological gapless boundary modes of topological insulators and superconductors \cite{SRFL08,Kitaev09,RSFL10,CTSR16}. 
In the presence of the coupling to the environment, however, we cannot retain these symmetries in their original form.
Causal fermionic theories require other onsite symmetries intrinsic to non-Hermitian systems \cite{LMC20, YPKH20}, which we call AZ$^\dagger$ symmetries \cite{KSUS19}. 
The AZ$^\dagger$ symmetries consist of 
TRS$^\dagger$: $H=U_T H^T U_T^\dagger$, $U_T U_T^*=\pm 1$,
PHS$^\dagger$: $H=-U_C H^* U_C^\dagger$, $U_C U_C^*=\pm 1$,
CS$^\dagger$: $H=-U_\Gamma H^\dagger U_\Gamma^\dagger$, $U_\Gamma U_\Gamma=1$,
%\begin{align}
%\mbox{TRS$^\dagger$}&: \quad H=U_T H^T U_T^\dagger, \quad &U_T U_T^*=\pm 1,
%\nonumber\\
%\mbox{PHS$^\dagger$}&: \quad H=-U_C H^* U_C^\dagger, \quad &U_C U_C^*=\pm 1,
%\nonumber\\
%\mbox{CS$^\dagger$}&:  \quad H=U_\Gamma H^\dagger U_\Gamma^\dagger, \quad &U_\Gamma U_\Gamma=1,
%\end{align}
where $H$ is the Hamiltonian, and $U_T$, $U_C$ and $U_\Gamma$ are unitary matrices \cite{KSUS19}. 
For a Hermitian $H$, 
the AZ$^\dagger$ symmetries coincide with the original AZ symmetries.
The presence and absence of the AZ$^\dagger$ symmetries define the ten-fold AZ$^\dagger$ classes \cite{KSUS19}. See Table \ref{TABLE1}.
One can easily check that the onsite decay constant term $-i\Gamma {\bm 1}$ due to dissipation respects the AZ$^\dagger$ symmetries \footnote{For details, see Sec.S3 in Supplement Material, which includes Ref. \cite{YPKH20}.}. 

A critical mathematical result for our theory is the extended Nielsen-Ninomiya theorem \cite{BS21}, which holds for systems in the AZ$^\dagger$ classes.
The theorem relates the bulk gapless points at the energies $E_\alpha$ with the topological charges $\nu_\alpha$ to the point-gap topological number $n$ at the reference energy $E_{\rm P}$, 
\begin{align}
n(E_{\rm P})=\sum_{{\rm Im}(E_\alpha-E_{\rm P})>0}\nu_\alpha=-\sum_{{\rm Im}(E_\alpha-E_{\rm P})<0}\nu_\alpha,        
\label{eq:extended_NN}
\end{align}
where the index $\alpha$ labels the gapless bulk states. 
This theorem implies that if topological gapless states have different lifetimes, 
there exists a region of $E_{\rm P}$ where $n$ is non-zero, namely we have a point-gap topological phase characterized by $n$.

Now we come back to our system, {\it i.e.} a $d$-dimensional topological insulator or superconductor with the open boundary condition in the $x_d$-direction.
Our system belongs to an AZ$^\dagger$ symmetry class when the system is coupled to the environment at $x_d=1$. 
Regarding the site index in the $x_d$-direction as an internal degree of freedom, we can identify the system as a $(d-1)$ dimensional system with "bulk" gapless states with the internal index $x_d=1, L_d$.  
Then, by the coupling to the environment at $x_d=1$, the $(d-1)$-dimensional bulk gapless states at $x_d=1$ have a different decay constant than those at $x_d=L_d$.
Therefore, from the extended Nielsen-Ninomiya theorem in Eq.(\ref{eq:extended_NN}), there exists a region of $E_{\rm P}$ where the $(d-1)$-dimensional point-gap topological number $n$ becomes non-zero.  
The non-zero value of $n$ is given by the $d$-dimensional bulk topological number of the original topological insulator/superconductor
since the total topological charge of the gapless states at $x_d=1$ coincides with it up to sign.

\begin{table}[t]
\centering
 \caption{Point-gap topological table for insulators and superconductors with dissipation at $x_d=1$.
Here $\delta=(d-1)-D$, where $d$ is the spatial dimension of the topological insulators and superconductors and $D$ is the dimension of a sphere surrounding a topological defect. The topological defect goes through the $x_d$-direction. The superscripts SE and BS indicate the topological numbers predicting non-Hermitian skin effects and boundary states, respectively. See also Ref.\cite{NBS22}. For $D=0$, this table reproduces that for $(d-1)$-dimensional point-gap phases in AZ$^\dagger$ classes in \cite{KSUS19}.}
 \begin{tabular}{ccccccc} \hline \hline
    AZ$^\dagger$ class & ~TRS$^\dagger$~ & ~PHS$^\dagger$~ & ~~CS~~ & ~$\delta=0$~ & ~$\delta=1$~ & ~$\delta=2$~ \\ \hline \hline
		\multirow{1}{*}{A}
		& $0$ & $0$ & $0$ & $0$ & $\mathbb{Z}^{\rm SE}$ & $0$ \\ 
		\multirow{1}{*}{AIII}
		& $0$ & $0$ & $1$ & $\mathbb{Z}$ &$0$ & $\mathbb{Z}^{\rm BS}$ \\ \hline 
    \multirow{1}{*}{$\text{AI}^{\dag}$}
    & $+1$ & $0$ & $0$ &$0$ & $0$ & $0$ \\ 
    \multirow{1}{*}{$\text{BDI}^{\dag}$}
    & $+1$ & $+1$ & $1$ &$\mathbb{Z}$ & $0$ & $0$ \\ 
    \multirow{1}{*}{$\text{D}^{\dag}$}
    & $0$ & $+1$ & $0$ & $\mathbb{Z}_2$  & $\mathbb{Z}^{\rm SE}$ & $0$ \\ 
    \multirow{1}{*}{$\text{DIII}^{\dag}$}
    & $-1$ & $+1$ & $1$ &\Zt & $\mathbb{Z}_2^{\rm SE}$ & $(2\mathbb{Z}+1)^{\rm SE}$ \\ 
       &  &  &  & &  & $2\mathbb{Z}^{\rm BS}$ \\ 
    \multirow{1}{*}{$\text{AII}^{\dag}$}
    & $-1$ & $0$ & $0$ & $0$ & $\mathbb{Z}_{2}^{\rm SE}$ & $\mathbb{Z}_2^{\rm SE}$\\ 
    \multirow{1}{*}{$\text{CII}^{\dag}$}
    & $-1$ & $-1$ & $1$ &2\Z & $0$ & $\mathbb{Z}_2^{\rm BS}$ \\ 
    \multirow{1}{*}{$\text{C}^{\dag}$}
    & $0$ & $-1$ & $0$ & $0$ & $2\mathbb{Z}^{\rm SE}$ & $0$ \\ 
    \multirow{1}{*}{$\text{CI}^{\dag}$}
    & $+1$ & $-1$ & $1$ &$0$ & $0$ & $2\mathbb{Z}^{\rm BS}$ \\ \hline \hline
  \end{tabular}
 \label{TABLE1} 
\end{table}

{\it Predictions.--}
The non-zero $(d-1)$-dimensional point-gap topological number $n$ gives rise to several consequences in the physical properties of the system.  
First,  it predicts the appearance of $(d-2)$-dimensional boundary modes or skin modes when imposing the additional open boundary condition on a different direction than $x_d$, say the $x_{d-1}$-direction \cite{NBS22}.
For $d=2$, the non-zero $n$ predicts a second-order non-Hermitian skin effect like the Chern insulator case in Fig.\ref{fig:QWZ}.
The second-order skin modes form a generalized Kramers pair \cite{SHEK12, KSUS19} when the original system has fermionic time-reversal symmetry. 
For $d=3$, an odd $n$ in class DIII$^\dagger$ (time-reversal invariant topological superconductor) and a non-trivial $n$ in class AII$^\dagger$ (topological insulator) imply the in-gap non-Hermitian skin effects. Still, other non-zero $n$s' predict boundary modes.
(See Table \ref{TABLE1} with $D=0$ introduced below.)

Second, the proposed system also may have similar localized modes in the presence of topological defects. 
The topological defects should go through the $x_d$-direction since our theory treats the site index in the $x_d$-direction as an internal degree of freedom.
Then, we can obtain the point-gap topological table in the presence of such topological defects by generalizing the argument by Teo and Kane \cite{TK10} to point-gap topological phases. See Table \ref{TABLE1}.
In the new table, the space dimension $d-1$ of the point-gap topological phases is replaced by $\delta=(d-1)-D$ $(\ge 0)$, where $D$ $(\le d-1)$ is the dimension of a sphere surrounding the topological defect. 
In particular, if we insert the $\pi$-flux ($D=1$) in the $x_3$-direction through a three-dimensional time-reversal invariant superconductor with an odd number of the three-dimensional winding number or a three-dimensional time-reversal invariant topological insulator, we have a non-trivial $\mathbb{Z}_2^{\rm SE}$ number for class DIII$^\dagger$ or class AII$^\dagger$ with $\delta=1$, in the presence of dissipation at $z=1$.
Thus, we obtain the non-Hermitian skin modes localized on the $\pi$-flux.

%{\it Non-Hermitian stabilization of topological boundary states.--}
%In addition to the phenomena above, 
Finally, the point-gap topology also stabilizes the original topological boundary modes of the topological insulator/superconductor.  
In general, the topological boundary modes of the topological insulator/superconductor have tiny gaps because of the mixing with those at an opposite boundary. 
Therefore, for finite $L_d$, they are not always gapless and do not have well-defined topological charges in a mathematically rigorous sense. 
In contrast, when the point-gap topological number $n$ becomes non-trivial, the extended Nielsen-Ninomiya theorem in Eq.(\ref{eq:extended_NN}) ensures the well-defined topological charges $\nu_\alpha$, which implies that the mixing disappear and the tiny gaps close.  
Indeed, the gapless modes at $x_d=1$ and those at $x_d=L_d$ have different imaginary parts of the energy, and thus they do not mix.
The gap closing of the boundary modes should be observed in high-resolution spectrum-sensitive experiments and sharpens the topological phenomena of the boundary modes.

{\it Examples.--}
We check the validity of our scheme in various topological materials.
For $d=1$, the dissipation effect for a superconducting nanowire has been discussed in literature \cite{PN13,SCPA16,APPSA19,OS21,LLWJX22}. 
The coupling of a Majorana end state to the environment is shown to give a non-trivial zero-dimensional point-gap $\mathbb{Z}_2$ number \cite{OS21}. 
It has been also demonstrated that the dissipation stabilizes the Majorana end state \cite{LLWJX22}. 
For $d=2$, we have already shown above that our theory for a Chern insulator gives the second-order non-Hermitian skin effect.
For $d=3$, exactly the same scheme was discussed for a three-dimensional time-reversal invariant topological insulator \cite{OS23}, which showed that non-Hermitian skin modes appear in the $\pi$-flux. (See Fig. 5b in Ref.\cite{OS23}.)
In Fig.\ref{fig:3dCSTI}, we also show the result for the three-dimensional chiral symmetric topological insulator, 
\begin{align}
H_{\rm CSTI}({\bm k})&= [\sin k_x + (1-\cos k_y)] \sigma_x \tau_x
+\sin k_y \sigma_y \tau_x
\nonumber\\
&+\sin k_z \sigma_z \tau_x + (-2+\sum_{i=x,y,z}\cos k_i)\sigma_0 \tau_y,   
\label{eq:3DCSTI}
\end{align}
where $\sigma_\mu$ and $\tau_\mu$ are the Pauli matrices, CS is $U_\Gamma=\sigma_0\tau_z$ and the onsite dissipation term $-i\Gamma \sigma_0\tau_0$ is placed at $z=1$.  
The system realizes the $\delta=2$ ($d=3$, $D=0$) AIII class with $\mathbb{Z}^{\rm BS}=1$. 
Under the open boundary conditions in both $z$ and $y$-directions, the system hosts a boundary state inside the point gap, as expected \footnote{See Sec.S1 in Supplement Material for spatial profiles of open boundary modes, which includes Refs. \cite{OS23,NONSS23}.}.

{\it Summary.--}
We propose a universal platform for point-gap topological phases constructed from topological insulators and superconductors. 
Using various independent arguments, we establish that dissipation on a boundary of $d$-dimensional topological materials results in $(d-1)$-dimensional point-gap topological phases.  
We also confirm the validity of our proposal for various topological materials.

Our scheme applies to any topological materials in the original topological periodic table \cite{SRFL08,Kitaev09,RSFL10,CTSR16}. For instance,  by connecting a metal to an edge of a quantum Hall state in graphene, we can realize a point-gap topological phase similar to Fig.\ref{fig:QWZ}. The resulting non-Hermitian skin effect can be observed as the chiral tunneling effect \cite{YY20}. Another candidate is a topological superconducting nanowire with a Zeeman field \cite{Sato-Ando-review17}. By coupling a lead to one of the ends of the nanowire, the system displays a point-gap topological phase in 0D class D$^\dagger$. We can also use the topological insulator Bi$_2$Se$_3$ and the variants \cite{Ando-review13}  to similarly realize a point-gap topological phase in 2D class AII$^\dagger$.

%In this paper, we have focused on topological materials classified by on-site fundamental symmetries and give general results.  However, crystalline symmetries also should provide additional topological structures in non-Hermitian systems.
%We hope to come back this problem in the future.

\vspace{1ex}
Note added: A part of the present work was reported in \cite{INS23}.
We are aware of related works \cite{XKXZS23,SGLK23} after the completion of this work.

\vspace{1ex}
{\it Acknowledgments.--}
This work was supported by JST CREST Grant No. JPMJCR19T2 and KAKENHI Grant No. JP20H00131.
D.N. was supported by JST, the establishment of university fellowships towards the creation of science technology innovation, Grant Number JPMJFS2123.
N.O. was supported by JSPS KAKENHI Grant No.~JP20K14373.

\bibliography{mainv2}

%%%%% Supplemental Material %%%%%%%%%%%%%%%%%%%%%
\widetext
\pagebreak

\renewcommand{\theequation}{S\arabic{equation}}
\renewcommand{\thefigure}{S\arabic{figure}}
\renewcommand{\thetable}{S\arabic{table}}
\renewcommand{\thesection}{S\arabic{section}}

\setcounter{equation}{0}
\setcounter{figure}{0}
\setcounter{table}{0}
\setcounter{section}{0}

\begin{center}
{\bf \large Supplemental Material}
\end{center}

\section{S1. Skin modes versus in-gap boundary modes}

Here we clarify the difference between skin modes and in-gap boundary modes. As discussed in \cite{OS23,NONSS23}, the right and left eigenstates for a skin mode should be localized in opposite boundaries, 
while those for an in-gap boundary mode are localized in the same boundary. 

\begin{figure}[h]
 \begin{center}
  \includegraphics[scale=0.13]{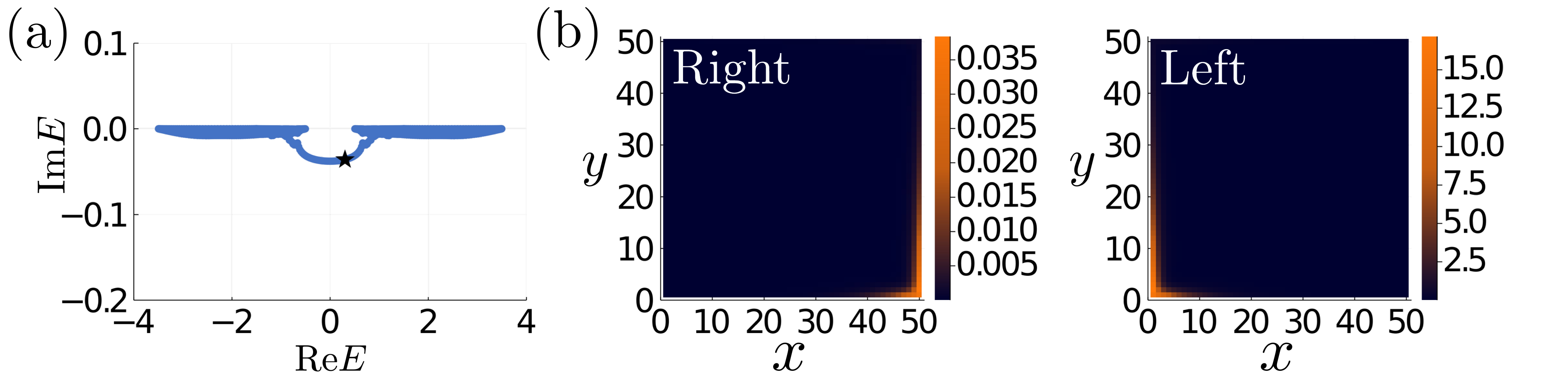}
  \caption{The QWZ model ($m=-1.5$, $L_x=L_y=50$) in Eq.(3) with $-i\Gamma \sigma_0$ term ($\Gamma=0.2$) at the $y=1$ boundary. (a) The complex spectrum under the open boundary condition in the $x$-direction. (b) The density profile of the right and left eigenstates for a corner skin mode with the energy $E=0.30-0.04 i$ (the star symbol in (a)).\label{2dclassA-SM}}
  \end{center}
\end{figure}

Figure \ref{2dclassA-SM} depicts the density profile of a corner skin mode in Fig.3 (d). As mentioned above, the right and left eigenstates of the corner skin mode are localized at different corners. In Fig.\ref{2dclassA-SM} (b), we impose the biorthogonal normalization on the corner skin mode where the inner product between the right and left eigenstate is 1.
Because the right and left eigenstates rarely overlap, the amplitude of the left eigenstate becomes large. (Here the inner product of the right eigenstate is normalized as 1.)

\begin{figure}[h]
 \begin{center}
  \includegraphics[scale=0.13]{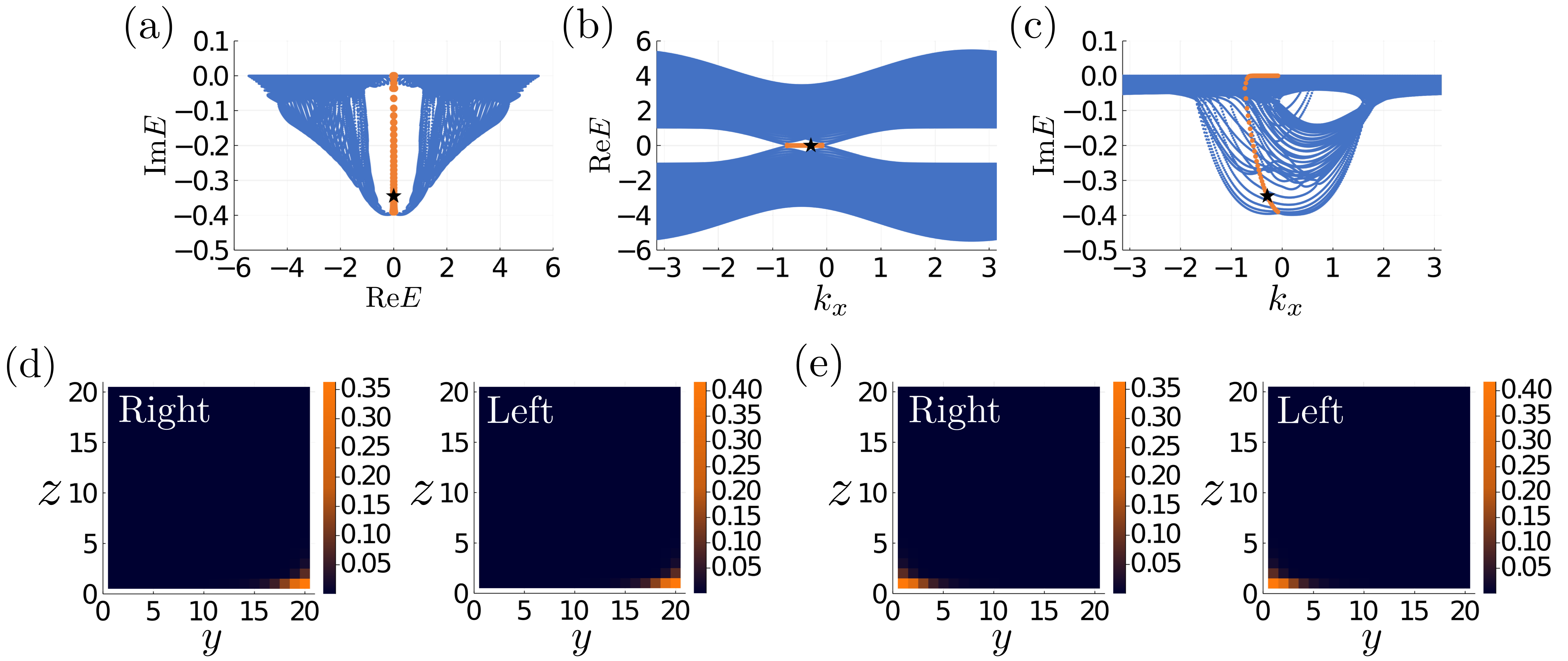}
  \caption{3D chiral symmetric topological insulator  ($L_x=300$, $L_y=L_z=20$) in Eq.(5) with 
  $-i\Gamma\sigma_0\tau_0$ ($\Gamma=0.2$) term at the $z=1$ boundary. (a,b,c) Complex spectrum of the system under the periodic boundary conditions in the $x$-direction and the open boundary condition in the $y$-directions. The orange curves indicate the in-gap boundary states. The in-gap boundary states are doubly degenerate because of mirror reflection symmetry in Eq.(\ref{Seq:mirror}). 
  (d) The density profile of the right and left eigenstates for an in-gap boundary state with the energy $E=-0.34 i$. (the star symbols in (a,b,c)). (e) The density profile of the right and left eigenstates for the mirror symmetry partner of the in-gap boundary state in (d). \label{3dclassAIII-SM}}
  \end{center}
\end{figure}

In contrast, the in-gap boundary modes in Fig.4 exhibit different behaviors.
As illustrated in Figs.\ref{3dclassAIII-SM} (d) and (e),
the right and left eigenstates of the in-gap boundary modes in Fig.4 are localized at the same corner. %\begin{figure}[btp]
Note that the three-dimensional chiral symmetric topological insulator $H_{\rm CSTI}({\bm k})$ in Eq.(5) has an additional mirror reflection symmetry,
\begin{align}
M_y H_{\rm CSTI}({\bm k}) M_y^\dagger=H(k_x,-k_y,k_z),
\quad M_y=\sigma_y\tau_y.
\label{Seq:mirror}
\end{align}
This symmetry preserves even when we impose the open boundary condition in the $z$-direction and place the on-site dissipation term $-i\Gamma\sigma_0\tau_0$ on the boundary at $z=1$. Because of this additional symmetry, the in-gap boundary modes have two-fold degeneracy.

\section{S2. Derivation of Dissipation term}

In this section, we microscopically derive the dissipation term used in the main text. 
As we shall show below, if we couple a good metal, whose bandwidth is much larger than that of the topological material, to the topological material, then we naturally obtain the constant dissipation term. (See Eq.(\ref{Seq:dissipation}).)

Let us consider a $d$ dimensional topological insulator described by the lattice Hamiltonian
\begin{align}
\hat{H}_{\rm top}=\sum_{ij}a^\dagger_{i}(H_{\rm top})_{ij}a_{j},    
\label{Seq:TI}
\end{align}
where $i$ is the index labeling the lattice site and other internal degrees of freedom such as spin and orbital, and $a_i$ and $a_i^\dagger$ are fermionic annihilation and creation operators.
As in the main text, we assume the open boundary condition in the $x_d$ direction and the periodic boundary conditions in the other directions.
Then, as an environment, we couple a metal described by the lattice Hamiltonian 
\begin{align}
\hat{H}_{\rm env}=\sum_{ij}b^\dagger_i(H_{\rm env})_{ij}b_j    
\end{align}
to the topological insulator at the bottom ($x_d=L$) with 
the coupling term
\begin{align}
\hat{H}_{\rm coupling}=\sum_{ij}a^\dagger_i (\tau)_{ij}b_j+{\rm h.c.},    
\end{align}
where $b_i$ and $b_i^\dagger$ are fermionic annihilation and creation operators.
The total Hamiltonian is $\hat{H}=\hat{H}_{\rm top}+\hat{H}_{\rm env}+\hat{H}_{\rm coipling}$, which is rewritten in the following simple form:
\begin{align}
\hat{H}=\sum_{\alpha i, \beta j}c^\dagger_{\alpha i}(H)_{\alpha i, \beta j}c_{\beta j}    
\end{align}
where $\alpha=1,2$ is the index distinguishing the topological insulator and the environment metal with $c_{1i}=a_i$, $c_{2i}=b_i$, $c^\dagger_{1i}=a^\dagger_i$, $c^\dagger_{2i}=b^\dagger_i$, 
and $H_{\alpha i,\beta j}$ is the matrix Hamiltonian given by
\begin{align}
H_{\alpha i,\beta j}=
\left(
\begin{array}{cc}
 (H_{\rm top})_{ij}    &  (\tau)_{ij}\\
 (\tau*)_{ji}    & (H_{\rm env})_{ij}
\end{array}
\right).
\end{align}
The Hamiltonians $\hat{H}_{\rm top}$ and $\hat{H}_{\rm env}$ are Hermitian and so is the total Hamiltonian $\hat{H}$.

To derive the effective Hamiltonian, we employ the retarded Green function defined by
\begin{align}
G^{\rm R}_{\alpha i, \beta j}(t)=-i\theta(t)\langle c_{\alpha i}(t)c_{\beta j}^\dagger
+c^\dagger_{\beta j}c_{\alpha i}(t)\rangle,   
\end{align}
where $c_{\alpha i}(t)=e^{i\hat{H}t}c_{\alpha i} e^{-i\hat{H}t}$ and $\langle \cdots\rangle={\rm tr}(e^{-\beta\hat{H} }\cdots)/{\rm tr}(e^{-\beta \hat{H}})$. 
Using the Heisenberg equation 
$i\partial_t c_{\alpha i}(t)=\sum_{\beta j}H_{\alpha i,\beta,j}c_{\beta j}$, we have
%\begin{align}
%\left(
%\begin{array}{cc}
%G_{\rm top} & G_{\rm te}\\
%G_{\rm et} & G_{\rm env}
%\end{array}
%\right)^{-1}=
%\left(
%\begin{array}{cc}
%E+i\eta-H_{\rm top}    &  -\tau\\
% -\tau^{\dagger}    & E+i\eta-H_{\rm env}
%\end{array}
%\right),
%\end{align}
%which satisfies 
\begin{align}
\left(
\begin{array}{cc}
E+i\eta-H_{\rm top}    &  -\tau\\
 -\tau^{\dagger}    & E+i\eta-H_{\rm env}
\end{array}
\right)
\left(
\begin{array}{cc}
G^{\rm R}_{\rm top} & G^{\rm R}_{\rm te}\\
G^{\rm R}_{\rm et} & G^{\rm R}_{\rm env}
\end{array}
\right)=1,
\end{align}
where $G^{\rm R}_{1i,1j}=(G^{\rm R}_{\rm top})_{ij}$, $G^{\rm R}_{1i,2j}=(G^{\rm R}_{\rm te})_{ij}$, $G^{\rm R}_{2i,1j}=(G^{\rm R}_{\rm et})_{ij}$ and $G^{\rm R}_{2i,2j}=(G^{\rm R}_{\rm env})_{ij}$. Here and below, the index $i$ is often implicit.
This equation leads to
\begin{align}
&(E+i\eta-H_{\rm top})G^{\rm R}_{\rm top}-\tau G^{\rm R}_{\rm et}=1,
\nonumber\\
&(E+i\eta-H_{\rm env})G^{\rm R}_{\rm et}-\tau^\dagger G^{\rm R}_{\rm top}=0,
\label{Seq:green}
\end{align}
then, introducing the retarded Green function $g_{\rm env}$ of the environment as
\begin{align}
g_{\rm env}^{-1}=(E+i\eta-H_{\rm env}),    
\end{align}
we obtain 
\begin{align}
G^{\rm R}_{\rm et}=g_{\rm env}\tau^\dagger G^{\rm R}_{\rm top}.    
\end{align}
from the second equation in Eq. (\ref{Seq:green}).
Then, the first equation in Eq.(\ref{Seq:green}) yields
\begin{align}
(E+i\eta-H_{\rm top}-\tau g_{\rm env}\tau^\dagger)G^{\rm R}_{\rm top}=1,    
\end{align}
which implies that the topological insulator coupled to the environment has the effective Hamiltonian \cite{Datta95}
\begin{align}
H_{\rm eff}=H_{\rm top}+\tau g_{\rm env} \tau^\dagger.    
\label{Seq:effective}
\end{align}
As we show immediately, the second term of the effective Hamiltonian is non-Hermitian and provides the dissipation term in the main text.

To evaluate the second term of Eq.(\ref{Seq:effective}), we assume a simple form of the coupling $\tau$,
\begin{align}
\tau_{ij}=\tau_{\rm c} \delta_{i_\perp, L_d}\delta_{j_\perp, L_d+1}\delta_{{\bm i}_{\parallel},{\bm j}_{\parallel}}
\delta_{\sigma\sigma'},
\end{align}
with a complex number $\tau_c$.
Here we explicitly write down the lattice site $(\vec{i}_\parallel, i_\perp)$ and the internal degrees of freedom $\sigma$ in the indices $i=({\bm i}_\parallel, i_\perp, \sigma)$ and $j=({\bm j}_\parallel, j_\perp, \sigma')$, and $i_\perp=L_d$ indicates the bottom boundary of the topological material. (The topological material extends between $i_\perp=1$ and $i_\perp=L_d$, and the metal does at $i_\perp\ge L_d+1$.)
This coupling leads to
\begin{align}
(\tau g_{\rm env} \tau^\dagger)_{ij}=
\left\{
\begin{array}{cl}
  |\tau_{\rm c}|^2 (g_{\rm env})_{({\bm i}_\parallel, L_d+1,\sigma) ({\bm j}_\parallel, L_d+1, \sigma')},  & \mbox{for $i_\perp=j_\perp=L_d$} \\
  0,   & \mbox{others}
\end{array}
\right.,
\end{align}
and thus, the second term is non-zero only at the bottom boundary of the topological material. 
For further evaluation, we use the lattice translation symmetry along the boundary, which enables the following momentum representation of $g_{\rm env}$,
\begin{align}
(g_{\rm env})_{({\bm i}_\parallel, L_d+1,\sigma) ({\bm j}_\parallel, L_d+1, \sigma')}=
\frac{1}{V_{d-1}}\sum_{{\bm k}_\parallel}g_{\rm env}({\bm k}_\parallel, L_d+1)\delta_{\sigma,\sigma'}e
^{i{\bm k}_\parallel ({\bm i}_\parallel-{\bm j}_\parallel)},  
\end{align}
where we assume that the metal Hamiltonian $H_{\rm emv}$ is independent of $\sigma$ for simplicity, and $V_{d-1}$ is the volume of the $(d-1)$-dimensional boundary. If we further assume that the metal is symmetric concerning the $\pi$-rotation normal to the boundary, we find that the imaginary part of 
$(\tau g_{\rm env}\tau^\dagger)_{ij}$ with $i_\perp=j_\perp=L_d$ is given by
\begin{align}
{\rm Im}(\tau g_{\rm env}\tau^\dagger)_{ij}
&=\frac{|\tau_c|^2}{V_{d-1}}\sum_{{\bm k}_\parallel}
{\rm Im}\left[
g_{\rm env}({\bm k}_\parallel, L_d+1)
\right]
\delta_{\sigma,\sigma'}e^{i{\bm k}_\parallel ({\bm i}_\parallel-{\bm j}_\parallel)}.   
\nonumber\\
&=-\frac{|\tau_c|^2}{\pi V_{d-1}}\sum_{{\bm k}_\parallel}\rho_{\rm env}({\bm k}_\parallel, E)
e^{i{\bm k}_\parallel ({\bm i}_\parallel-{\bm j}_\parallel)},   
\end{align}
where $\rho_{\rm env}({\bm k}_\parallel, E)$ is the density of states of the metal with parallel momentum ${\bm k}_\parallel$ and energy $E$ at the boundary $i_\parallel=L_d+1$. 
Here we have used the Green function representation of the density of states:
\begin{align}
\rho_{\rm env}({\bm k}_\parallel, E)=-\pi{\rm Im}\left[
g_{\rm env}({\bm k}_\parallel, L_d+1)
\right].
\end{align}
For a good metal of which the bandwidth is much larger than that of the topological insulator, we can approximate the density of state by the constant averaged value $\bar{\rho}_{\rm env}$ \cite{ATEKOW14},  
then we obtain
\begin{align}
{\rm Im}(\tau g_{\rm env}\tau^\dagger)_{ij}=-\frac{|\tau_c|^2}{\pi}\bar{\rho}_{\rm env}\delta_{{\bm i}_\parallel, {\bm j}_\parallel}\delta_{i_\perp, L_d}\delta_{j_\perp, L_d}\delta_{\sigma,\sigma'},  
\label{Seq:dissipation}
\end{align}
which provide the dissipation term $-i \Gamma$ with $\Gamma=|\tau_c|^2\bar{\rho}_{\rm env}/\pi$ at the bottom boundary of the topological insulator.
Note that ${\rm Re}[g_{\rm env}({\bm k}_\parallel, L_d+1)]$ only renormalize $H_{\rm top}$ at the boundary, and can be ignored if the bulk gap of the topological insulator is large enough. 

If we consider a topological superconductor instead of a topological insulator, the Hamiltonian in Eq.(\ref{Seq:TI}) is modified as
\begin{align}
\hat{H}_{\rm top}= \sum_{ij}a^\dagger_i (H_{\rm normal})_{ij}a_j+ \frac{1}{2}\sum_{ij}\left[
a_i^\dagger (\Delta)_{ij}a^\dagger_j+{\rm h.c} \right].  
\end{align}
Still, we can derive the dissipation term in a similar manner by using the Green function in Nambu space.
From a parallel argument, we have the dissipation term at the boundary 
\begin{align}
\begin{pmatrix}
-i\Gamma \delta_{{\bm i}_\parallel, {\bm j}_\parallel}\delta_{i_\perp, L_d}\delta_{j_\perp, L_d}\delta_{\sigma,\sigma'}
& 0\\    
0 & 
-i\Gamma \delta_{{\bm i}_\parallel, {\bm j}_\parallel}\delta_{i_\perp, L_d}\delta_{j_\perp, L_d}\delta_{\sigma,\sigma'}
\end{pmatrix},    
\end{align}
with $\Gamma=|\tau_c|^2\bar{\rho}_{\rm env}/\pi$ in the Nambu space representaion.

\section{S3. AZ$^\dagger$ symmetry}

In this section, we show that AZ symmetry in the total Hamiltonian results in AZ$^\dagger$ symmetry in the effective Hamiltonian in Sec. S2.

First, we introduce AZ symmetry in terms of fermionic operators. The AZ symmetry is local symmetry consisting of TRS, PHS, and CS defined by
\begin{align}
\mbox{TRS}:
\quad 
&\hat{T}c^\dagger_{\alpha i} \hat{T}^{-1}=\sum_{j}c^\dagger_{\alpha j}(U^\alpha_T)_{ji},
\quad 
\hat{T}c_{\alpha i} \hat{T}^{-1}=\sum_{j}c_{\alpha j}(U^\alpha_T)^*_{ji},
\\
\mbox{PHS}:
\quad
&\hat{C}c^\dagger_{\alpha i} \hat{C}^{-1}=\sum_{j}c_{\alpha j}(U^\alpha_C)^*_{ji},
\quad
\hat{C}c_{\alpha i} \hat{C}^{-1}=\sum_{j}c^\dagger_{\alpha j}(U^\alpha_C)_{ji},
\\
\mbox{CS}:
\quad
&\hat{\Sigma}c^\dagger_{\alpha i} \hat{\Sigma}^{-1}=\sum_{j}c_{\alpha j}(\Gamma^\alpha)^*_{ji},
\quad
\hat{\Sigma}c_{\alpha i} \hat{\Sigma}^{-1}=\sum_{j}c^\dagger_{\alpha j}(\Gamma^\alpha)_{ji},
\end{align}
where $\alpha$ is the label distinguishing the topological material ($\alpha=1$) and the environment metal ($\alpha=2$),  $\hat{T}$ and $\hat{\Sigma}$ are anti-unitary operators, $\hat{C}$ is a unitary operator, and $U^\alpha_T$, $U^\alpha_C$, and $\Gamma^\alpha$ are unitary matrices.
In the absence of PHS, $c_{\alpha i}$ is the electron operator ($c_{1i}=a_{i}$ and $c_{2i}=b_{i}$), but in the presence of PHS, it represents the corresponding Nambu spinor.
Since AZ symmetry is local symmetry, it does not mix different $\alpha$.

For the quadratic Hamiltonian
\begin{align}
\hat{H}=\sum_{\alpha i,\beta j}c^\dagger_{\alpha i}(H)_{\alpha i,\beta j}c_{\beta j},    
\end{align}
the above operations reproduce the standard AZ symmetry:
\begin{align}
\mbox{TRS}: \quad &[\hat{H}, \hat{T}]=0 \quad \Leftrightarrow 
\quad  (U_T^\alpha)_{ip}(H)^*_{\alpha p, \beta q} (U_T^{\beta\dagger})_{q j}=(H)_{\alpha i, \beta j},
\\
\mbox{PHS}: \quad  &[\hat{H},\hat{C}]=0 \quad \Leftrightarrow 
\quad  (U_C^\alpha)_{ip}(H^T)_{\alpha p, \beta q} (U_C^{\beta\dagger})_{q j}=-(H)_{\alpha i, \beta j},
\\
\mbox{CS}: \quad &[\hat{H}, \hat{\Sigma}]=0  \quad \Leftrightarrow 
\quad   (\Gamma^\alpha)_{ip}(H^\dagger)_{\alpha p, \beta q} (\Gamma^{\beta\dagger})_{q j}=-(H)_{\alpha i, \beta j},
\quad {\rm tr}H^*=0.
\label{Seq:CS2}
\end{align}
Note that one can omit ${\rm tr}H^*=0$  in Eq.(\ref{Seq:CS2}) if $\hat{H}$ is Hermitian as one can derive it from the first equation in the right-hand side of Eq.(\ref{Seq:CS2}).

To examine AZ symmetry in the effective Hamiltonian, let us consider the retarded Green function for the topological material \cite{YPKH20}:
\begin{align}
(G^{\rm R}_{\rm top})_{ij}=-i\theta(t)\langle a_i(t)a^\dagger_j+a^\dagger_ja_i(t)\rangle.
\end{align} 
First, we examine TRS in the effective Hamiltonian. When $\hat{H}$ has TRS, {\it i.e.} $\hat{T}^{-1}\hat{H}\hat{T}=\hat{H}$, 
we have
\begin{align}
\langle a_i(t)a^\dagger_j\rangle&={\rm tr}(e^{-\beta\hat{T}^{-1}\hat{H}\hat{T}}e^{i\hat{T}^{-1}\hat{H}\hat{T}t}a_i e^{-i\hat{T}^{-1}\hat{H}\hat{T} t}a^\dagger_j)/{\rm tr}(e^{\beta\hat{H}})
\nonumber\\
&={\rm tr}(\hat{T}^{-1}e^{-\beta\hat{H}}e^{-i\hat{H}t}\hat{T}a_i\hat{T}^{-1} e^{i\hat{H}t}\hat{T}a^\dagger_j\hat{T}^{-1}\hat{T}). 
\label{Seq:aa}
\end{align}
Since $\hat{T}$ is anti-unitary, we have
\begin{align}
\langle n|\hat{T}^{-1}\hat{O}|m\rangle&=(\langle \hat{T} n|\hat{T}\hat{T}^{-1}\hat{O}|m\rangle)^*
\nonumber\\
&=(\langle \hat{T} n|\hat{T}\hat{T}^{-1}\hat{O}\hat{T}^{-1}\hat{T}|m\rangle)^*
\nonumber\\
&=(\langle \hat{T} n|\hat{O}\hat{T}^{-1}|\hat{T}m\rangle)^*,
\end{align}
which leads to ${\rm tr}(\hat{T}^{-1}\hat{O})={\rm tr}(\hat{O}\hat{T}^{-1})^*$. 
From this, Eq.(\ref{Seq:aa}) is recast into 
\begin{align}
\langle a_i(t)a^\dagger_j\rangle&=
[{\rm tr}(e^{-\beta\hat{H}}e^{-i\hat{H}t}\hat{T}a_i\hat{T}^{-1} e^{i\hat{H}t}\hat{T}a^\dagger_j\hat{T}^{-1})]^*/{\rm tr}(e^{-\beta\hat{H}})
\nonumber\\
&=\left[
\sum_{kl}
{\rm tr}(e^{-\beta\hat{H}}e^{-i\hat{H}t}a_k e^{i\hat{H}t}a^\dagger_l)(U_T^1)^*_{ki} (U_T^1)_{lj}
\right]^*/{\rm tr}(e^{-\beta\hat{H}})
\nonumber\\
&=\sum_{kl}(U_T^1)_{ki}
{\rm tr}(a_l e^{-i\hat{H}t}a_k^\dagger e^{i\hat{H}t}e^{-\beta\hat{H}})
(U_T^1)^*_{lj}/{\rm tr}(e^{-\beta\hat{H}})
\nonumber\\
&=\sum_{kl}(U_T^1)_{ki}
{\rm tr}(e^{-\beta\hat{H}} a_l(t)a_k^\dagger)
(U_T^1)^*_{lj}/{\rm tr}(e^{-\beta\hat{H}})
\nonumber\\
&=\sum_{kl}(U_T^1)^\dagger_{jl}
\langle a_l(t)a_k^\dagger\rangle
(U_T^1)_{ki},
\end{align}
where we have used the Hermiticity of $\hat{H}$.
Thus, we have 
\begin{align}
(G^{\rm R}_{\rm top})_{ij}&=(U_T^1)^\dagger_{jl}\left[-i\theta(t)\langle a_l(t) a_k^\dagger+a_k^\dagger a_l(t)\rangle\right](U_T^1)_{ki}
\nonumber\\
&=(U_T^1)^\dagger_{jl}(G^{\rm R}_{\rm top})_{lk}(U_T^1)_{ki}, 
\end{align}
which leads to the transpose version of TRS, which we call TRS$^\dagger$, in the effective Hamiltonian for the topological material.
\begin{align}
U_T^1 H_{\rm eff}^T (U_T^1)^\dagger=H_{\rm eff}.
\end{align}

In a similar manner,  we can show that PHS and CS of the total Hamiltonian $\hat{H}$  leads to the complex conjugation version of PHS, 
\begin{align}
U_C^1 H_{\rm eff}^* (U_C^1)^\dagger=-H_{\rm eff},
\end{align}
which we call PHS$^\dagger$, and CS
\begin{align}
\Gamma^1  H_{\rm eff}^\dagger (\Gamma^1)^\dagger=-H_{\rm eff},
\end{align}
respectively, in the effective Hamiltonian.
Therefore, AZ symmetry of the total Hamiltonian results in AZ$^\dagger$ symmetry (TRS$^\dagger$, PHS$^\dagger$ and CS ) in the effective Hamiltonian \cite{YPKH20}. 

\section{S4. Relation between the Hatano-Nelson model and the chiral edge state with dissipation.}

In this section, we explain the relation between the Hatano-Nelson model and the chiral edge state with dissipation.
 
First, we derive Eq.(1) from Eq.(2) for $|\Delta|\ll 1$. 
In Eq.(1), we consider a gapless mode at ${\rm Re}E=0$ in the complex energy $E$ plane, and thus we focus on the spectrum of Eq.(2) around ${\rm Re}E=0$.
Since the Hatano-Nelson model $h_{\rm HN}(k_x)=-v\sin k_x+i(\Gamma/2)\cos k_x-i(\Gamma/2)$
satisfies ${\rm Re}h_{\rm HN}(k_x)=0$ at $k_x=0,\pi$, 
we consider the spectrum of Eq.(2) around $k_x=0,\pi$.

The eigenvalues of Eq.(2) are
\begin{align}
E_\pm(k_x)=\frac{1}{2}\left[vk_x-v\sin k_x+i(\Gamma/2)\cos k_x-i(\Gamma/2)\pm
\sqrt{\left(vk_x+v\sin k_x-i(\Gamma/2)\cos k_x +i(\Gamma/2)\right)^2+4|\Delta|^2}\right].    
\end{align}
Around $k_x=0$, these eigenvalues are $E_\pm (k_x)=\pm \sqrt{(v k_x)^2+|\Delta|^2}+\cdots$, and thus they are gapful at ${\rm Re}E=0$. 
On the other hand, around $k_x=\pi$, they give
\begin{align}
E_{+}(k_x)=v\pi +O(|\Delta|^2),
\quad
E_{-}(k_x)=v(k_x-\pi)-i\Gamma+O(|\Delta|^2),
\end{align}
and thus,  $E_-(k_y)$ becomes gapless at ${\rm Re}E=0$. Then, by shifting $k_x$ by $\pi$, $E_-(k_y)$ gives Eq.(1).
For small $|\Delta|$, Eq.(2) has a non-zero winding number because the Hatano-Nelson model has the non-zero winding number, implying the non-trivial topology in Eq.(1).

\begin{figure}[h]
 \begin{center}
  \includegraphics[scale=0.15]{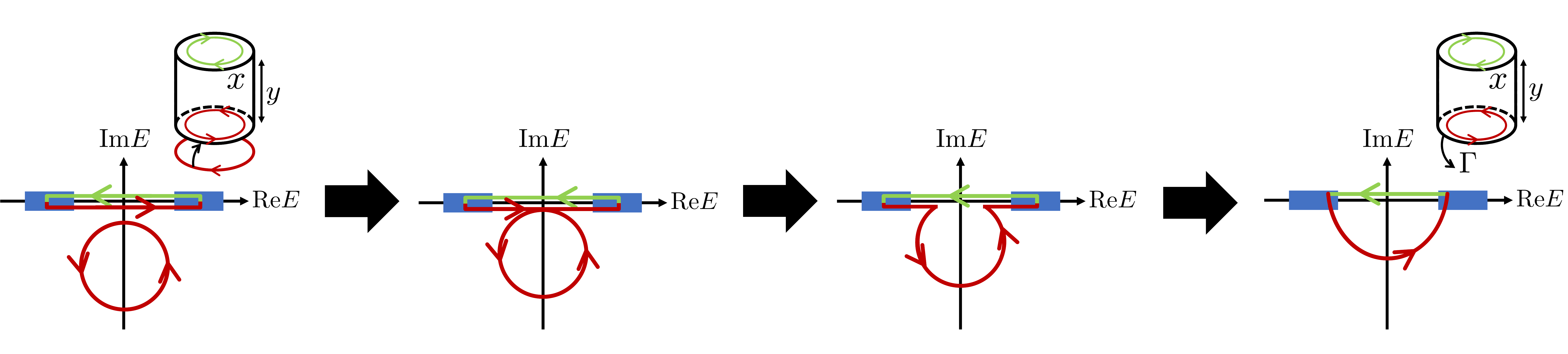}
  \caption{The Hatano-Nelson model attached to a Chern insulator at the bottom boundary (leftmost). The chiral edge mode at the bottom boundary merges with the Hatano-Nelson model (middles), and it eventually reproduces the spectrum of the chiral edge mode with the imaginary part of the energy (rightmost).
  \label{Sfig:HN_chiral}}
  \end{center}
\end{figure}

We also illustrate how the Hatano-Nelson model gives the dissipation term in the chiral edge mode in Fig.\ref{Sfig:HN_chiral}.
The leftmost panel in Fig.\ref{Sfig:HN_chiral} shows the spectrum (blue, green, and red lines) of a Chern insulator under the open boundary condition in the $y$-direction together with the complex spectrum of the Hatano-Nelson model (red circle). 
By adiabatically increasing the coupling between the Hatano-Nelson model and the bottom boundary of the Chern insulator, the whole spectrum changes from the left to the right in Fig.\ref{Sfig:HN_chiral}. Finally, we have the complex spectrum of the chiral edge mode with the dissipation in the rightmost of Fig.\ref{Sfig:HN_chiral}. 

\begin{figure}[h]
 \begin{center}
  \includegraphics[scale=0.18]{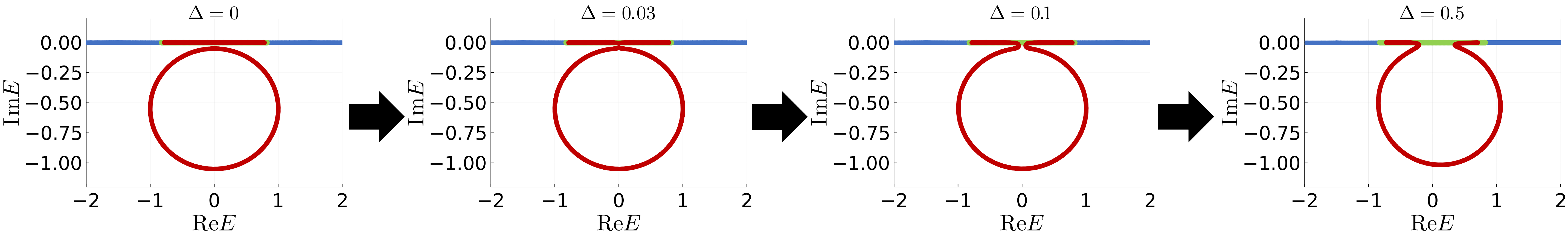}
  \caption{A merging process between the Hatano-Nelson model and the chiral edge mode of the QWZ model ($L_y=50,m=-1.5,v=\Gamma=1$).\label{Sfig:NH+QWZ}}
  \end{center}
\end{figure}

Figure \ref{Sfig:NH+QWZ} shows the same adiabatical process obtained by numerical calculation.
Here we consider the model $H = H_{\text{QWZ}} + H_{\text{HN}} + H_{\Delta}$
with
\begin{align}
    &H_{\rm QWZ}=\left(\dfrac{1}{2}\right)\Biggl\{\sum_{i_{y}=1}^{L_y}\sin k_x[c^{\dagger}_{i_{y}\uparrow}(k_x)c_{i_{y}\downarrow}(k_x)+c^{\dagger}_{i_{y}\downarrow}(k_x)c_{i_{y}\uparrow}(k_x)] + [-c^{\dagger}_{i_{y}+1\uparrow}(k_x)c_{i_{y}\downarrow}(k_x)+c^{\dagger}_{i_{y}+1\downarrow}(k_x)c_{i_{y}\uparrow}(k_x)] \nonumber \\
    &\qquad\quad + (m+\cos k_x)[c^{\dagger}_{i_{y}\uparrow}(k_x)c_{i_{y}\uparrow}(k_x)-c^{\dagger}_{i_{y}\downarrow}(k_x)c_{i_{y}\downarrow}(k_x)]
    + [c^{\dagger}_{i_{y}+1\uparrow}(k_x)c_{i_{y}\uparrow}(k_x)-c^{\dagger}_{i_{y}+1\downarrow}(k_x)c_{i_{y}\downarrow}(k_x)]\Biggr\} + \text{h.c.},
    \nonumber\\
    &H_{\text{HN}}= [-v\sin k_x+i(\Gamma/2)\cos k_x-i(\Gamma/2)]c^{\dagger}_{0\uparrow}(k_x)c_{0\uparrow}(k_x)
    -(0.05i)c^{\dagger}_{0\uparrow}(k_x)c_{0\uparrow}(k_x) 
    + (10i) c^{\dagger}_{0\downarrow}(k_x)c_{0\downarrow}(k_x),\nonumber\\
    &H_{\Delta} = \Delta c^{\dagger}_{1\uparrow}(k_x)c_{0\uparrow}(k_x) + \Delta c^{\dagger}_{1\downarrow}(k_x)c_{0\downarrow}(k_x) + \text{h.c.},
\end{align}
where $H_{\rm QWZ}$ is the (second quantized) Qi-Wu-Zhang model, $H_{\rm HN}$ is the Hatano-Nelson model, $H_{\Delta}$ is the coupling between them, and $c_{i_y\sigma}(k_x)$ and $c^\dagger_{i_y\sigma}(k_x)$ are the creation and annihilation operators of electron with spin $\sigma$ and the momentum $k_x$ in the $x$ direction at site $i_y$ in the $y$-direction. The numerical result reproduces the adiabatic deformation in Fig.\ref{Sfig:HN_chiral}.

\end{document}